\documentstyle[prl,aps,epsfig]{revtex}

\begin{document}
\draft
\title{Evidence of Vortices on the Insulating Side of the Superconductor-Insulator
Transition}
\author{N. Markovi\'{c}, A. M. Mack, G. Martinez-Arizala,C. Christiansen and A. M.
Goldman}
\address{School of Physics and Astronomy, University of Minnesota, Minneapolis,\\
MN 55455, USA}
\date{\today}
\maketitle

\begin{abstract}
The magnetoresistance of ultrathin insulating films of Bi has been studied
with magnetic fields applied parallel and perpendicular to the plane of the
sample. Deep in the strongly localized regime, the magnetoresistance is
negative and independent of field orientation. As film thicknesses increase,
the magnetoresistance becomes positive, and a difference between values
measured in perpendicular and parallel fields appears, which is a linear
function of the magnetic field and is positive. This is not consistent with
the quantum interference picture. We suggest that it is due to vortices
present on the insulating side of the superconductor-insulator transition.
\end{abstract}


\narrowtext
\twocolumn
In the limit of zero temperature, ultrathin films of metals become either
insulating or superconducting depending on the strength of the disorder \cite
{Haviland} or the applied magnetic field \cite{Hebard}. The
superconductor-insulator (SI) transition has been described by a boson
Hubbard \cite{Fisher} model which also predicts a metallic phase for films
on the margin between insulating and superconducting behavior whose
resistance is ''universal.'' Within this model, at $T=0$, the
superconducting state is a Cooper pair condensate with localized vortices,
and the insulating state is a vortex condensate with localized Cooper pairs.
There has been no full generalization to nonzero temperature. The boson
Hubbard model has been challenged by the results of tunneling experiments 
\cite{Valles} carried out on thin films of metals which have been
interpreted as showing that the amplitude of the order parameter is
extremely small or zero at the SI transition and in the insulating state. On
the other hand, Hall and longitudinal resistance studies on indium oxide
films \cite{Ruel}, driven out of the superconducting state by magnetic
fields, were interpreted as evidence of two different insulating phases, one
presumably a Bose insulator of localized Cooper pairs and the other the
usual Fermi insulator of localized electrons. If this interpretation were
correct, films insulating by virtue of disorder might behave similarly.
Furthermore, if there were vortices and Cooper pairs on the insulating side
of the transition, then one might expect to see some evidence of vortex
motion in the magnetoresistance of such films.

The magnetoresistance (MR) of disordered two dimensional electronic systems 
\cite{Lee} in both strongly (SL) and weakly (WL) localized regimes is found
to be dominated by two basic mechanisms: an orbital effect \cite
{Nguyen,Sivan} which is due to quantum interference between electron paths,
and a spin effect \cite{Eto,Maekawa} which arises from Zeeman splitting of
the electron spin states. The former is flux driven, and therefore highly
sensitive to whether the magnetic field is applied in the direction
perpendicular to the plane of the sample or parallel to it, while the latter
is isotropic and depends only on the magnetic field strength. In systems
with strong spin-orbit (SO) coupling, both mechanisms predict a negative MR
in the SL regime and a positive MR in the WL regime. The characteristic
anisotropy due to the orbital effect, as well as the signature of the spin
effect, has been observed in a number of experiments \cite{Ovadyahu,Giordano}%
. Here we report the results of MR measurements carried out on ultrathin
insulating films of Bi close to the SI transition in magnetic fields both
parallel and perpendicular to the plane of the sample. The observed
anisotropic component of the MR is a linear function of the magnetic field,
always positive and it increases monotonically through the SL-WL crossover.
This is not consistent with it being an orbital effect. We suggest that this
contribution arises from the flow of vortices, which may offer new evidence
in favor of the boson Hubbard model.

The ultrathin Bi films used in this study were evaporated on top of a 10\AA\
thick layer of amorphous Ge, which was pre-deposited onto the substrate
which was a 0.75mm thick single-crystal of $SrTiO_{3}$ $(100)$. The
substrate temperature was kept below 20 K during all depositions and all the
films were grown $in$ $situ$ under UHV conditions ($\sim 10^{-10}$Torr).
Under such circumstances, successive depositions could be carried out
without contamination to increase the film thickness in increments of
0.25\AA\ up to 15\AA . Film thicknesses were determined using a previously
calibrated quartz crystal monitor. Films prepared in this manner are
believed to be homogeneous, since it has been found that they become
connected at an average thickness on the order of one monolayer \cite
{Strongin}. Resistance measurements were carried out using a standard dc
four-probe technique with current bias up to 50nA. Current-voltage
characteristics were linear up to at least 1$\mu $A. Magnetic fields up to
20kG parallel to the plane of the sample, and up to 12kG perpendicular to
the plane of the sample were applied using a superconducting split-coil
magnet.

The temperature dependence of the conductance is logarithmic at higher
temperatures, as expected for a weakly localized 2D system, while at the
lowest temperatures it becomes activated and could be fit by $%
G=G_{0}exp(-[T_{0}/T]^{x})$ where x = 1/2. At intermediate temperatures,
there is a crossover region between the logarithmic and exponential
behavior, which moves towards lower temperatures in thicker films.

Fig. \ref{MR vs B} shows the MR as a function of magnetic field at 0.7K for
the four most insulating films in both parallel and perpendicular field. The
thinnest film, sample 1, shows a negative MR with a fractional change of up
to 12\% in a 20kG field, which seems to be field-orientation independent.
Although not shown on the graph, additional data taken at different
temperatures confirm this isotropic behavior. Sample 2, which is 0.25\AA\
thicker, also shows a negative MR, but the fractional change is now less
than 4\% in a 20kG field. For this sample, the parallel MR is greater than
the perpendicular. This anisotropy becomes more obvious for sample 3, which
shows a small positive MR for both directions of the field. The MR of sample
4 is even more positive and the difference between the parallel and the
perpendicular field increases further.

The MR changes sign from negative to positive in both parallel and
perpendicular magnetic field at the temperatures and the thicknesses for
which the system is at the crossover between the SL and the WL regime
(crossover being loosely defined as the region where the behavior of the
conductance changes from activated to diffusive). A similar sign change has
been observed in thin films of Au, Ag and Cu by Hsu et al. \cite{Hsu}, where
the MR was considered to be dominated by the orbital effect.

In our case, the MR is dominated by a spin effect. One can conclude that
from the fact that plotting the longitudinal MR against $\mu _{B}B/k_{B}T$
makes all the data collapse on a single curve. This also implies that the
orbital contribution is negligible in parallel field. In the variable range
hopping regime, a mechanism proposed by Eto \cite{Eto} predicts an
isotropic, negative MR due to Zeeman effect even in the presence of SO
interactions. A model by Maekawa and Fukuyama \cite{Maekawa}, also based on
Zeeman effect, predicts a positive MR in the WL regime. Our results are in
qualitative agreement with these pictures.

We define the anisotropic component of the MR as the difference between the
transverse and the longitudinal magnetoresistances $R_{a}=(\Delta R_{\perp
}-\Delta R_{\parallel })/R(0)$ \cite{Jing}. Fig. \ref{MR vs d} shows that $%
R_{a}$ is a linear function of the magnetic field for ten different
thicknesses{\bf .} We will argue that this anisotropy cannot be due to any
standard orbital effect. In the SL regime, in which conduction occurs
through variable range hopping, the orbital MR can be described by a model
originally proposed by Nguyen, Spivak and Shklovskii (NSS) \cite{Nguyen} in
which an applied magnetic field increases the conductance. The critical
percolation path approach \cite{Sivan} yields a fractional change in the
resistance $\Delta R/R(0)\propto B^{2}$ at low fields, which increases with
decreasing temperature. A similar behavior is predicted in the presence of
SO scattering, although the magnitude of the MR is expected to decrease as
the SO scattering increases \cite{Meir,Medina}.

In a 2D system, the sample thickness is smaller than the hopping length,
which forces all hops to be in the plane of the sample. This restriction on
hop orientations leads to a MR anisotropy and a MR\ in a perpendicular field
which should be much larger than in a parallel field. Even though the MR of
sample 2 is slightly anisotropic, this anisotropy cannot be due to the
orbital mechanism described above as the magnitude of the MR is smaller in
perpendicular than in parallel fields. In addition, a predicted quadratic
behavior with magnetic field was not observed.

In the WL regime, the orbital mechanism is different. In systems with strong
SO coupling there is a positive MR, which becomes larger with decreasing
temperature as the electron phase coherence time increases. Quantitatively,
the theory predicts $\Delta R/R(0)\propto B^{2}$ in the low field limit and $%
\Delta R/R(0)\propto lnB$ in the high field limit \cite{Lee,Bergmann}. In a
2D system, a field parallel to the plane will make no contribution to the
flux through the time reversed loops, and no parallel MR is expected.
Samples 3 and 4 considered in Fig. \ref{MR vs B} are in the WL regime where
the conductance shows a logarithmic dependence on temperature. The MR is
positive and its magnitude is larger in perpendicular field, which might be
expected from the model described above, but further analysis reveals some
inconsistencies. Namely, it was not possible to obtain a satisfactory fit to
the WL orbital effect theory \cite{Bergmann}, which predicts $R_{a}\propto
B^{2}$ in the low field. Instead, $R_{a}$ is clearly a linear function of
the magnetic field for a wide range of thicknesses and temperatures.
Furthermore, the anisotropy due to the orbital effect is expected to change
sign \cite{Hsu} at the SL-WL crossover. The anisotropic response first
occurs at the temperatures and thicknesses at which the films are in the SL
regime, and it persists into the WL regime, but its magnitude is always
positive and it changes monotonically through the crossover, as shown on
Fig. \ref{crossover}. All this suggests that standard orbital effects cannot
be the mechanism responsible for the anisotropy.

Orbital MR has been observed in a number of experiments on several other
systems \cite{Ovadyahu,Giordano} and is often used to obtain various
scattering times. One might therefore ask why we do not see any orbital
effect? It is known that orbital MR is suppressed in the presence of strong
SO coupling \cite{Meir,Medina}, which might explain why the contribution to
the MR from the orbital effect is so small compared to indium-oxide films.
It is possible that this contribution would become overwhelmed by the
effects of superconducting ordering before becoming significant (this would
not be the case for Au, Ag and Cu films, which can get deep into the WL
regime where the orbital effect is substantial, without becoming
superconducting). Indeed, closer to the SI transition, $R_{a}$ starts to
deviate slightly from the linear field dependence (the very top of the Fig. 
\ref{MR vs d}), acquiring a small quadratic term. This might be due to the
orbital effect, or possibly Maki-Thompson (MT) effect, which has the same
field dependence as the WL orbital effect \cite{Larkin}.

If the boson-Hubbard model actually described the insulating state near the
SI transition, then the insulator might be able to sustain point-like
vortices because of the nonvanishing Cooper pair density. If these vortices
were to move freely \cite{Wallin} in response to currents, they might
produce a flux flow contribution to the MR of the system. This would result
in $R_{\perp }$-$R_{\parallel }$ always being positive and proportional to
the magnetic field \cite{Tinkham}, which is the observed behavior.
Increasing the thickness (decreasing the disorder) and lowering the
temperature drives the system deeper into this putative Bose insulator
state, (the most insulating films do not exhibit this effect and are Fermi
insulators) where Cooper pairs and vortices are more likely to form. The
anisotropy due to flux flow in this regime would therefore become more
pronounced with increasing thickness and decreasing temperature, as observed
in the measurements. In other words, the linear response, except for its
relatively small magnitude, very much resembles the MR due to flux flow in
superconducting films, which our samples become when made just slightly
thicker. If the process were a dynamical effect, i.e., the vortex-like MR
response was occurring in the presence of some kind of order parameter
amplitude and phase fluctuations, this picture might not be incompatible
with the interpretation of tunneling spectra on the insulating side of the
transition, that the superconducting gap and the average order parameter
amplitude are both zero \cite{Valles}. Alternatively, and consistent with
the original boson Hubbard picture, a nonvanishing superconducting order
parameter amplitude could persist into the insulating regime, even though
superconductivity is destroyed by phase fluctuations. Phase fluctuations can
reduce the gap \cite{Ferrel}, as they are pair breaking \cite{Lemberger}.
This can lead to gapless superconductivity. It is also possible that the
local gap is destroyed by disorder, but the fermionic degrees of freedom are
highly suppressed and maybe even dynamically irrelevant \cite{Wallin}. In
support of this view are indications of superconducting effects in films of $%
In_{2}O_{3}$ driven into the insulating state by magnetic fields, in the
work of Paalanen et al. \cite{Ruel} and in insulating films of $In_{2}O_{3}$
in the work of Gantmakher and co-workers \cite{Gantmakher}.

Further experimental work and a theoretical model suitable for the
intermediate disorder at finite temperatures are needed to resolve this
issue. In particular, we are attempting to detect the presence of vortices
in the insulating regime by searching for vortex shot noise in a manner
similar to the study of Knoedler and Voss \cite{Knoedler} carried out on
granular aluminum films in the superconducting state.

We gratefully acknowledge useful discussions with L. Glazman and A.
Finkel'stein. This work was supported in part by the National Science
Foundation under Grant No. NSF/DMR-9623477.

\begin{figure}[tbp]
\caption{Magnetoresistance as a function of magnetic field for: a) four
films of different nominal thicknesses: 10.50\AA\ (circles), 10.75\AA\
(diamonds), 11.00\AA\ (triangles) and 11.25\AA\ (squares); b) sample 3
(d=11.00\AA ) at two different temperature: 0.7K (circles) and 0.3K
(triangles). Full symbols represent the perpendicular field, and open
symbols represent the parallel field.}
\label{MR vs B}
\end{figure}

\begin{figure}[tbp]
\caption{The difference between the perpendicular and parallel
magnetoresistance R$_{a}$ as a function of magnetic field for ten films of
different thicknesses at 0.7K. Full lines are linear fits. The zeros of the
vertical scale are offset for clarity.}
\label{MR vs d}
\end{figure}

\begin{figure}[tbp]
\caption{The slope of R$_{a}$ vs B, which can be thought of as a measure of
the anisotropy, as a function of film thickness for six different films at
T=0.7K. Inset: The slope of R$_{a}$ vs B as a function of temperature for
sample 3 (d=11.00\AA ). The lines are guides to the eye. The arrows show
where the system is at the SL-WL crossover.}
\label{crossover}
\end{figure}


\begin{references}
\bibitem{Haviland}  D. B. Haviland, Y. Liu and A. M. Goldman, Phys. Rev.
Lett. {\bf 62}, 2180 (1989); Y. Liu, K. A. McGreer, B. Nease, D. B.
Haviland, G. Martinez, J. W. Halley and A. M. Goldman, Phys. Rev. Lett. {\bf %
67}, 2068 (1991); Y. Liu, D. B. Haviland, B. Nease and A. M. Goldman, Phys.
Rev. B {\bf 47}, 5931 (1993).

\bibitem{Hebard}  A. F. Hebard and M. A. Paalanen, Phys. Rev. Lett. {\bf 65}%
, 927 (1990); Ali Yazdani and Aharon Kapitulnik, Phys. Rev. Lett. {\bf 74},
3037 (1995).

\bibitem{Fisher}  M. P. A. Fisher, Phys. Rev. Lett. {\bf 65}, 923 (1990);
M.-C. Cha, M. P. A. Fisher, S. M. Girvin, M. Wallin and A. P. Young, Phys.
Rev. B {\bf 44}, 6883 (1991).

\bibitem{Valles}  J. M. Valles, Jr., R. C. Dynes and J. P. Garno, Phys. Rev.
Lett. {\bf 69}, 3567 (1992); S-Y. Hsu, J. A. Chervenak and J. M. Valles,
Jr., Phys. Rev. Lett. {\bf 75}, 132 (1995).

\bibitem{Ruel}  M. A. Paalanen, A. F. Hebard and R. R. Ruel, Phys. Rev.
Lett. {\bf 69}, 1604 (1992).

\bibitem{Lee}  For a review, see P. A. Lee and T. V. Ramakrishnan, Rev. Mod.
Phys. {\bf 57}, 287 (1985); S. Hikami, A. I. Larkin, and Y. Nagaoka, Prog.
Theor. Phys.{\bf \ 63}, 707 (1980).

\bibitem{Nguyen}  V. I. Nguyen, B. Z. Spivak and B. I. Shklovskii, JETP
Lett. {\bf 41}, 42 (1985); Sov.Phys. JETP {\bf 62}, 1021 (1985).

\bibitem{Sivan}  U. Sivan, O. Entin Wohlman and Y. Imry, Phys. Rev. Lett. 
{\bf 60}, 1566 (1988).

\bibitem{Eto}  M. Eto, Phys. Rev. B {\bf 51}, 13066 (1995).

\bibitem{Maekawa}  S. Maekawa and H. Fukuyama, J. Phys. Soc. Jpn. {\bf 50},
2516 (1981).

\bibitem{Ovadyahu}  D. Kowal, M. Ben-Chorin and Z. Ovadyahu, Phys. Rev. B 
{\bf 44}, 9080 (1991); A. Vaknin, A. Frydman, Z. Ovadyahu and M. Pollak,
Phys. Rev. B {\bf 54}, 13604 (1996); A. Frydman and Z. Ovadyahu, Solid State
Commun. {\bf 94}, 745 (1995).

\bibitem{Giordano}  N. Giordano and M. A. Pennington, Phys. Rev. B {\bf 47},
9693 (1993); F. Komori, S. Kobayashi and W. Sasaki, J. Phys. Soc. Japan {\bf %
52}, 36 (1983).

\bibitem{Strongin}  M. Strongin, R. S. Thompson, O. F. Kammerer and J. E.
Crow, Phys. Rev. B {\bf 1}, 1078 (1970).

\bibitem{Hsu}  S.-Y. Hsu and J. M. Valles, Jr., Phys. Rev. Lett. {\bf 74},
2331 (1995).

\bibitem{Jing}  T. W. Jing, N. P. Ong, T. V. Ramakrishnan, J. M. Tarascon
and K. Remschnig, Phys. Rev. Lett. {\bf 67}, 761 (1991).

\bibitem{Meir}  Y. Meir, N. S. Wingreen, O. Entin-Wohlman, and B. L.
Altshuler, Phys. Rev. Lett. {\bf 66}, 1517 (1991).

\bibitem{Medina}  E. Medina and M. Kardar, Phys. Rev. Lett. {\bf 66}, 3187
(1991); E. Medina, M. Kardar and R. Rangel, Phys. Rev. B {\bf 53}, 7663
(1996).

\bibitem{Bergmann}  G. Bergmann, Phys. Rep. {\bf 107}, 1 (1984); G.
Bergmann, Phys. Rev. Lett. {\bf 48}, 1046 (1982).

\bibitem{Larkin}  A. I. Larkin, Pis'ma Zh. Eksp. Teor. Fiz. {\bf 31}, 239
(1980) [JETP Lett. {\bf 31}, 219 (1980)].

\bibitem{Wallin}  M. Wallin, E. S. S\o rensen, S. M. Girvin and A. P. Young,
Phys. Rev. B {\bf 49}, 12115 (1994).

\bibitem{Tinkham}  M. Tinkham,''Introduction to Superconductivity'',
(McGraw-Hill, NewYork, 1975).

\bibitem{Ferrel}  R. A. Ferrel and C. J. Lobb (private communication).

\bibitem{Lemberger}  Soon-Gul Lee and Thomas R. Lemberger, Phys. Rev. B {\bf %
37}, 7911 (1988).

\bibitem{Gantmakher}  V. F. Gantmakher, M. V. Golubkov, J. G. S. Lok and A.
K. Geim, Zh. Exp. Teor. Fiz. {\bf 109}, 1765 (1996) [Sov. Phys. JETP {\bf 82}%
, 951 (1996)].

\bibitem{Knoedler}  C. M. Knoedler and R. F. Voss, Phys. Rev. B {\bf 26},
449 (1982).
\end{references}
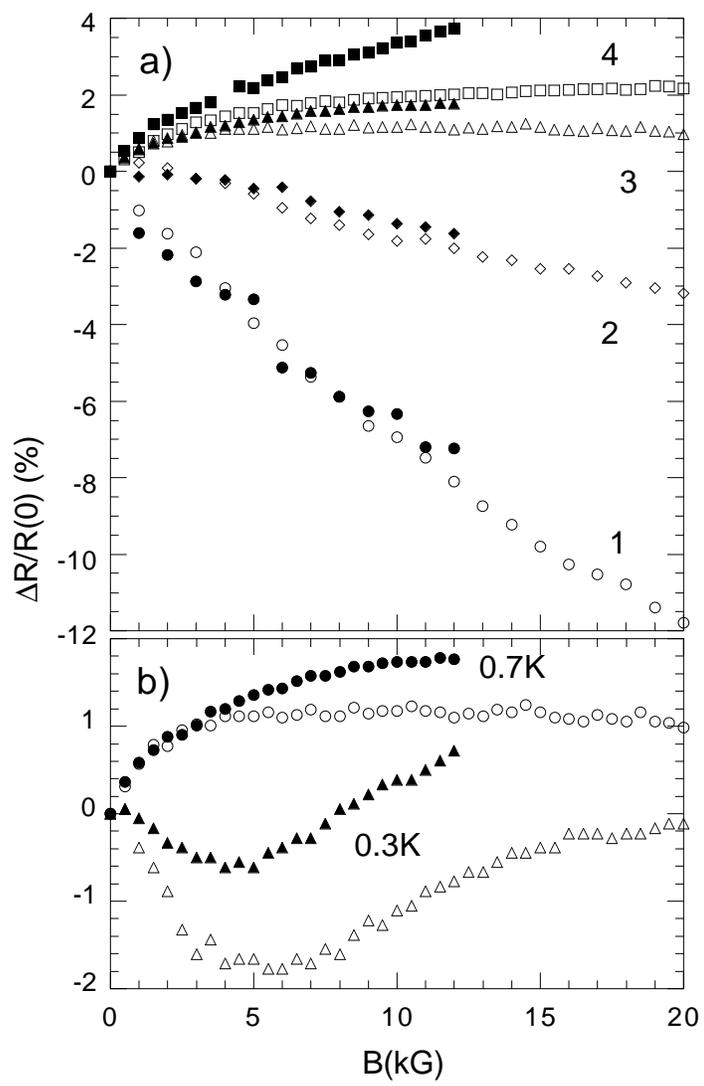
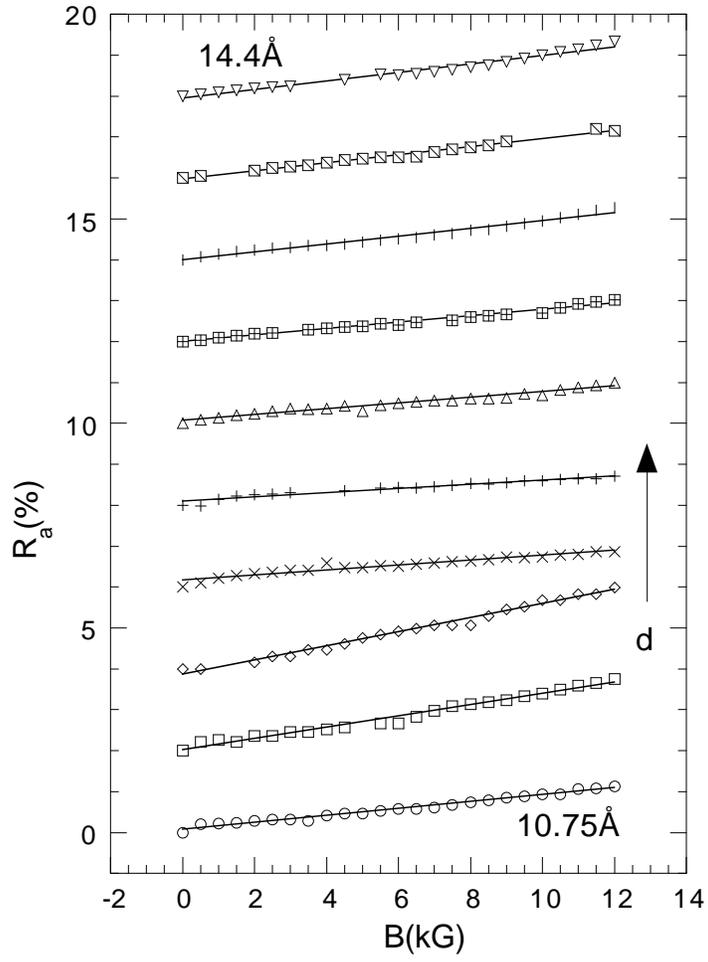
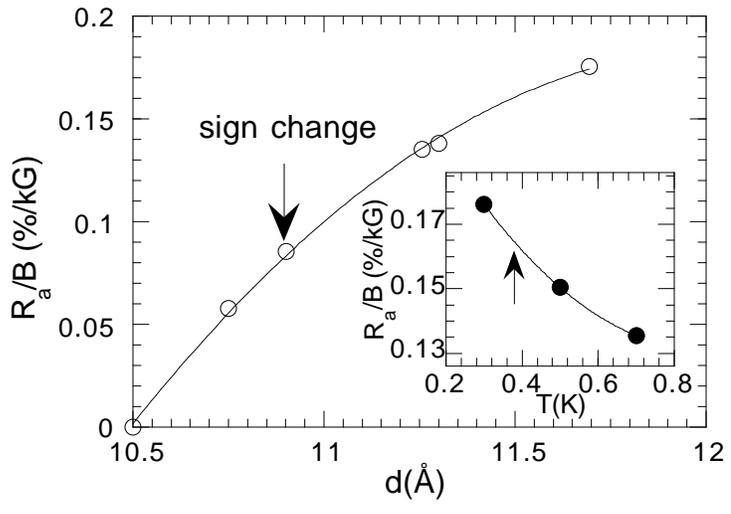
\end{document}